\documentclass[11pt]{article}
\usepackage{setspace}
\usepackage{amssymb}
\usepackage{amsmath}
\usepackage{amsfonts}

 \def\G{\Gamma} %\mbox{\boldmath $A$}

\def\be{\begin{equation}}

\def\ee{\end{equation}}

\newcommand\Ga{\Gamma}

\def\dd{\partial}

\def\bea{\begin{eqnarray}}

\def\eea{\end{eqnarray}}

\makeatletter
\def\blfootnote{\xdef\@thefnmark{}\@footnotetext}
\makeatother

\begin{document}

\singlespace

\begin{flushright} BRX TH-6319 \\
CALT-TH 2017-025
\end{flushright}

\vspace*{.3in}

\begin{center}

{\Large\bf Higher curvature gravities, unlike GR, cannot be bootstrapped from their (usual) linearizations}

{\large S.\ Deser}

{\it 
Walter Burke Institute for Theoretical Physics, \\
California Institute of Technology, Pasadena, CA 91125; \\
Physics Department,  Brandeis University, Waltham, MA 02454 \\
{\tt deser@brandeis.edu}
}
\end{center}

\begin{abstract}
We show that higher curvature order gravities, in particular the propagating quadratic curvature models, cannot be derived by self-coupling from their linear, flat space, forms, except through an unphysical version of linearization; only GR can. Separately, we comment on an early version of the self-coupling bootstrap.
\end{abstract}

\section{Introduction}
One, physically illuminating, way of understanding General Relativity (GR) is as a free spin $2$ $m=0$ field in flat space that must become self-coupled if it is to interact consistently with matter stress tensor sources. That is, unlike the $s=1$ Maxwell theory, that remains neutral when coupled to charges --- but like its Yang-Mills generalization 
--- GR is intrinsically nonlinear because it must be self-interacting, i.e., ``charged". This was shown [1] in closed form using its first order, Palatini, formulation. Further, the requirement is applicable starting from any, not just flat, background, thereby allowing for a cosmological term [2] as well. The interesting question was raised recently  whether other models, notably actions including those with quadratic curvature actions, could also be bootstrapped [3]. This note shows that they can, but only  at the cost of an (over?)extended definition of the linear starting point. If instead, we insist on the usual physical meaning of linearization, we show explicitly that quadratic gravities, the only other models with kinetic terms at flat space, do not bootstrap. Cubic and higher curvature terms (including Lanzos-Lovelock) are rather to be considered as sources and also cannot consistently bootstrap; of course Bianchi identity consistency requires that they be covariantized to be permitted.

\section{The GR bootstrap}
We begin with a brief resume of the relevant parts of [1]. The free flat space $s=2$ $m=0$ action has 
the schematic form
\begin{equation}
A_L[\eta, \Gamma , h] = \int \left[ \sqrt{-\eta} \eta^{\mu\nu} \hbox{Ric}^{(2)}_{\mu\nu} + \frak{h}^{\mu\nu} \hbox{Ric}^{(1)}_{\mu\nu}\right].
\end{equation}
Here Ric$^{(1)}$ is the part of the [covariant] Ricci tensor linear in $\Gamma$, while Ric$^{(2)}$ is the quadratic part; the full Ricci tensor is their sum.  No indices are needed since there is no possible ambiguity.  Above, $\eta$ is the flat metric, $\frak{h}$ is a contravariant tensor density, while $\Gamma$ is the affinity/connection treated as an independent variable. Thus, the stress tensor (or density) $T_{\mu\nu}$ of the system (1) becomes, in this notation, just Ric$^{(2)}(\Gamma)$, requiring a cubic self-coupling addition to the action: 
\begin{equation}
A_{NL} = \int (\frak{g}_0 + \frak{h}) \left[ \hbox{Ric}^{(2)}(\Gamma) + \hbox{Ric}^{(1)}(\Gamma) \right] = \int \frak{g} \hbox{Ric}(\Gamma).
\end{equation}
Here $\frak{g}_0^{\mu\nu}$ is the contravariant density $\sqrt{-\eta} \eta^{\mu\nu}$.  But now, only the combination $\frak g$, rather than the separate tensors $\frak{g}_0$ and $\frak{h}$ appears, and the bootstrap stops --- as it should, because (2) is the full Palatini GR action!  Actually,  we never had to specify whether $\Ga$ was an independent or (linearized) metric connection: read either way, (2) and its derivation are correct, when the above Palatini method is slightly generalized. To see this, consider the second order form of (1), where $\Ga$ is now taken to be $\G_L(h) \sim \eta^{..} \dd h$, so that  (1) becomes
\begin{equation}
A_L[\eta,h] = \int\sqrt{-\eta} \dd h [\eta \eta \eta]^{..} \dd h
\end{equation}
The $\eta \eta \eta$ form denotes a rank $6$ tensor cubic in the contravariant $\eta$s with appropriate index combinations. To begin the bootstrap in this form amounts to adding the corresponding stress-tensor coupling, namely
adding a term $\sim  \int h^{\mu\nu} \delta A_L/\delta \eta^{\mu\nu}$ to the linear action. [Note that here $h^{\mu\nu}$ is {\it not} $\frak{h}^{\mu\nu}$ but instead merely $\eta^{\mu\alpha} \eta ^{\nu\beta} h_{\alpha\beta}$]. That is, the second order bootstrap does {\it not} stop, but goes on indefinitely until the true inverse of $g_{\mu\nu}=(\eta_{\mu\nu}+h_{\mu\nu})$ --- and automatically $\frak{g}^{\mu\nu} =(\eta^{\mu\nu}+\frak{h}^{\mu\nu})$ --- is reached. 
 This is why Palatini form is simply cubic in its variables, while traditional second order form of GR is an infinite series in the metric (and its first derivatives), whether co- or contra-variant of any density order:\begin{equation}
\begin{aligned}
A_{NL}[g = \eta+h] &= \int\frak{g}^{..} \Ga(g) \Ga(g) \\
&= \int \sqrt{-g} (\dd_\alpha g_{\mu\nu}) [ g^{\alpha\beta} g^{\mu\kappa} g^{\nu\lambda} - g^{\alpha \beta} g^{\mu\nu} g^{\kappa \lambda} - 2 g^{\alpha \lambda} g^{\beta\nu} g^{\mu\kappa} + 2 g^{\mu\nu} g^{\alpha \kappa} g^{\beta\lambda}] (\dd_\beta g_{\kappa\lambda}),
\end{aligned}
\end{equation}
in its full index glory.  Had we included a (flat) volume,  $\int\sqrt{-\eta}$, term, it would have bootstrapped to the usual cosmological  $\int\sqrt{-g}$. Finally, to dispel one last possible cloud, we have minimized formalism by making the $h$``$T$" addition to $\eta$``$T$" according to the co/contra and density character of the initial $\eta$.
The quotes simply remind us that the tensor character of ``T" is determined by that of the $\eta$ chosen to start the bootstrap. Still, one might worry that the true Belinfante-Rosenfeld (say) ``T" is, instead, the contravariant tensor density coefficient of the covariant tensor $\eta$ in the action, so that this use of ``T" is cheating. Of course, the beauty of tensor notation is that it precludes cheating: Concretely, the chosen point of departure does not alter the outcome.   But let's verify this pedantically: we can express any other desired initial $\eta$ form as an (infinite series) in the one we have used. Then bootstrap each term with the stress tensor form corresponding to the desired $\eta$ to reach \ldots exactly the shortcut answer, as a moment's reflection shows --- same infinite series, but now in that $(\eta+h)$, rather than just in $\eta$ --- now re-sum: Trees saved!

Equation (4) is perhaps the most compact form of the EH action, one that shows that GR simply modifies the free field kinematics by a (metric) form factor. Indeed, a recent reformulation of GR [4] arrived at similar forms by demanding this property, also newly exploited in [5].  So one can fairly summarize GR --- and, as we shall see, only GR --- as ``self-coupling is general covariance".
[As a point of history, this form of the GR action has actually been known for (a least) eight decades [6], as the one containing only first metric derivatives, $A\sim \int \frak g^{\mu\nu} \Ga \Ga$; it is also, when written out in terms of the $3+1$, first order form variables, GR's ADM version.]

\section{Higher order Gravities}
  Only gravity actions quadratic in the curvature have flat space linearizations, unlike say $R^3$ models that start as $\int R_L^3(h)\sim \int h^3$.  The general $D=4$ quadratic curvature action is, owing to the Gauss-Bonnet identity, the two-parameter $A = \int [a R_{\mu\nu}^2 + b R^2]$. As is well-known (and obvious), Palatini and second-order formulations here describe very different models, whether or not we keep the GR term. The Palatini forms are easy to bootstrap, but at the cost of being (even) less physical than their more popular second order variants. There are actually two interpretations of ``Palatini" here, the obvious one treated here and a ``nonlinear linear" one treated in Appendix 1.  Since the GR and higher curvature parts bootstrap separately, we need treat only the latter. Their linearized actions are sums of the form 
 \begin{equation}
 A_L(\eta,h) = \int\sqrt{-\eta} R_{\mu\nu}^L(\Ga) (\eta \eta)^{\mu\nu\alpha\beta} R_{\alpha \beta}^L(\Gamma), \, \, \, \, \, \, \, \, \, \, \, 
 R_{\mu\nu}(\Ga) \sim \dd \Gamma ; 
 \end{equation}
unlike in full GR, we must drop the quadratic, $\Ga \Ga$, terms in $R_{\mu\nu}$ as they would lead to cubic and quartic ones in the action. Hence, in 2nd order, since $\Ga_L(h) \sim \dd h$, then $R_{\mu\nu}^L\sim \dd^2 h$ and the form corresponding to (3) is
\begin{equation}
A_L[\eta,h] = \int\sqrt{-\eta} \dd^2 h\,  [\eta \eta \eta \eta]^{..} \, \dd^2 h,
\end{equation}
where, as in (3), the $8$-index (this time) quartic in $\eta$'s indices are suitably distributed. Now we press the magic button, $\eta \rightarrow \eta+h$, most conveniently in terms of the contravariant tensor form of $\eta$ and $h$ (though as we saw, it doesn't matter in principle), to self-couple and obtain the analog of (4),
\begin{equation}
A_{NL}[g=\eta+h] = \int\sqrt{-g} \dd^2 g_{..} [g g g g]^{..} \dd^2 g_{..} .
\end{equation}
This is where thing go wrong, however: varying the metric in (7) manifestly cannot recover the $\Ga \Ga \Ga \Ga$ terms present in the $RR$ parts of the field equations; self-coupling is no longer invariance,
and (7) is not  $\int RR$. The problem obviously lies in the quadratic, rather than quartic, $h$-dependence 
at linear level, (6), traceable in turn to the $R_L^2$ in (5). It takes the more liberal definition of linearization of Appendix 1 to achieve this. One might argue that there is an ambiguity in
(6), absent in (3), in the placing of derivatives on $h_{..}$ prior to covariantizing, but there is a canonical
ordering in the linear action $\int R_L R_L \sim \int \dd^2 h_{..} \eta^{..} \dd^2 h_{..}$ that is absent (and indeed irrelevant) in the GR case.

A perhaps illuminating way to physically understand the failure of the bootstrap in the quadratic case is to note first that the linearized starting point is in fact the difference of two independent second-order actions, one of them in terms of a new set of --- Ostrogradskii --- variables. Each of these actions can now be bootstrapped independently, leading to two separate non-linear (second-order) universes, But that model does {\it not} represent the desired normal nonlinear quadratic gravity, where of course there is but one metric: the two original ones mix beyond linear order. 

\section*{Acknowledgements}
This work was supported by grants NSF PHY-1266107 and DOE\#desc0011632. I thank T. Ortin
for correspondence, J. Franklin for excellent translation, and Chris Fewster for useful suggestions.

\section*{Appendix 1. Bootstrap does work, BUT only with an (over?)extended definition of linearity.}

We carry out the alternative definition of linearization mentioned in text; it works because it treats the
--- in this Appendix, Palatini-like, independent $\Ga$ --- as ``matter" variables, to be kept to all orders: only the metric is to be initially linearized, while $R_{\mu\nu}(\G)$ maintains the form $\sim \dd \Ga +\Ga\Ga$, so that the ``linear" action is 
%\begin{equation}\label{A1}
\begin{equation}\label{A1}
A_L(\G; \eta) = \int\sqrt{-\eta} [\dd \G + \G \G] (\eta \eta) [\dd \Ga + \Ga \Ga] .
\end{equation}
We add a Lagrange multiplier term to enforce (linearized) metricity of the affinity $\Ga$, 
\begin{equation}\label{A2}
\Delta A_L = \int \lambda[ \G - \G_L(\eta,h)].
\end{equation}
This $\eta$-dependence must be bootstrapped --- here most conveniently in terms of the contravariant $\eta^{\mu\nu}$, via $\Ga_L\sim \eta \dd h \rightarrow  \G_{NL}\sim(\eta+h)^{..} \dd(\eta+h)_{..}$ to reach the full nonlinear affinity, 
\begin{equation}\label{A3}
\Delta A_{NL} = \int \lambda[\Ga - \Ga_{NL}(g = \eta + h)],
\end{equation}
along with the obvious required bootstrapping of the explicit $\eta$ in (\ref{A1}) to $\eta+h$.  [To see the $\Ga$ bootstrap in detail, write $\Ga^\alpha_{L \, \, \beta\gamma}= 1/2 \eta^{\alpha \delta} [-\dd_\delta h_\beta\gamma + \hbox{perm}]$ and replace $\eta^{..}$ by $(\eta+h)^{..}=g^{..}$ while adding $\dd \eta$ to $\dd h$.] But this is exactly the desired nonlinear result, that, using the $\lambda$ constraint $\Ga=\Ga(g)$, the final action is the sum of invariant terms in the full curvatures,
\begin{equation}\label{A4}
A_{NL}(g) = \int\sqrt{-g} R_{\mu\nu} (g g)^{\mu\nu\alpha\beta} R_{\alpha\beta}, \, \, \, \, \, \, \, \, \, \, \, R_{\mu\nu} = R_{\mu\nu}(\Ga(g)).
\end{equation}
This method manifestly covers any invariant, not just quadratic, powers of Riem, Ricci and $R$. What we have not covered are functionals involving (covariant) derivatives of the $R$s, because these terms require their own bootstrap, a straightforward, but separate, task. However, this ``linearization" definition --- treating the ``linearized" gravity variables $\Ga$ as if they were matter ones, kept to all orders --- seems less physical than the usual prescription of Sec 3.

\section*{Appendix 2. Another derivation of GR?}
Long before [1], the first  attempt at a bootstrap derivation of GR was given in [7], using a different formal technique to obtain ``$\eta \rightarrow \eta + h$" from self-coupling, leading to the requirement that the action must depend only on the latter combination. This raises the question whether it might have been used here as well. A re-reading of [7], however, prompts some queries. Firstly, the linear, 
$\sim \Box h$, term in the field equation was obtained as if it were part of a stress tensor, from varying an action $\sim \int h \hbox{Ricci}(\eta)$, a somewhat puzzling choice. Secondly it was explicitly assumed that the final action, because it only depended on $g$, was necessarily a coordinate  invariant, which is of course a stretch. Furthermore, it was stated that the resulting action would represent ANY invariant, whereas of course the 2nd order kinetic term is only compatible with GR, and not any other models! Nevertheless, the underlying idea that self-coupling leads to a functional that depends only on $(\eta+h)$ is correct; it was used --- in a different context --- in [2], for example.

\end{document}